\documentclass[aps, prl, reprint, superscriptaddress,  floatfix]{revtex4-1}
\usepackage{latexsym}
\usepackage{amsthm}
\usepackage{amssymb}
\usepackage{amsmath}
\usepackage[all]{xy}
\usepackage{float}
\usepackage{graphicx}
\usepackage{subfigure}
\usepackage{dcolumn}
\usepackage{bm}
\usepackage{bbm}
\usepackage[usenames, dvipsnames]{xcolor}
\usepackage{amsfonts}
\usepackage{cases}
\usepackage{multirow}
\usepackage{bigstrut}

\usepackage[pagebackref=false, colorlinks=true,
linkcolor=blue, citecolor=blue,urlcolor=blue,]{hyperref}

\begin{document}

\title{Constraints on Velocity and Spin Dependent Exotic
	Interaction at the Millimeter Scale with a Diamagnetic-levitated Force Sensor}

\author{Kenan Tian}
\affiliation{National Laboratory of Solid State Microstructures and Department of Physics,  Nanjing University, Nanjing 210093,
	China}

\author{Yuanji Sheng}
\affiliation{National Laboratory of Solid State Microstructures and Department of Physics, Nanjing University, Nanjing 210093,
	China}

\author{Rui Li}
\affiliation{National Laboratory of Solid State Microstructures and Department of Physics, Nanjing University, Nanjing 210093,
	China}

\author{Lei Wang}
\affiliation{National Laboratory of Solid State Microstructures and Department of Physics, Nanjing University, Nanjing 210093,
	China}

\author{Peiran Yin}
\email{ypr@nju.edu.cn}
\affiliation{National Laboratory of Solid State Microstructures and Department of Physics, Nanjing University, Nanjing 210093,
	China}
 
 \author{Shaochun Lin}
 \affiliation{CAS Key Laboratory of Microscale Magnetic Resonance and School of Physical Sciences, University of Science and Technology of China, Hefei 230026, China}
 
 \author{Dingjiang Long}
 \affiliation{National Laboratory of Solid State Microstructures and Department of Physics,  Nanjing University, Nanjing 210093,
	China}
 
 \author{Chang-Kui Duan}
  \affiliation{CAS Key Laboratory of Microscale Magnetic Resonance and School of Physical Sciences, University of Science and Technology of China, Hefei 230026, China}
\affiliation{Hefei National Laboratory, University of Science and Technology of China, Hefei 230088, China }

\author{Xi Kong}
\affiliation{National Laboratory of Solid State Microstructures and Department of Physics, Nanjing University, Nanjing 210093,
	China}
 
 \author{Pu Huang}
 \email{hp@nju.edu.cn}
\affiliation{National Laboratory of Solid State Microstructures and Department of Physics, Nanjing University, Nanjing 210093,
	China}

\author{Jiangfeng Du}
\email{djf@ustc.edu.cn}
\affiliation{CAS Key Laboratory of Microscale Magnetic Resonance and School of Physical Sciences, University of Science and Technology of China, Hefei 230026, China}
\affiliation{Hefei National Laboratory, University of Science and Technology of China, Hefei 230088, China }
\affiliation{Anhui Province Key Laboratory of Scientific Instrument Development and Application, University of Science and Technology of China, Hefei 230026, China}

\affiliation{Institute of Quantum Sensing and School of Physics, Zhejiang University, Hangzhou 310027, China}

\begin{abstract}
 Light bosons, beyond the standard model and as prominent candidates for dark matter, can mediate velocity and spin dependent exotic interaction between electron spins and nucleons. At short ranges, it remains an open challenge to test this exotic interaction with high precision. Here, we present a method based on diamagnetic-levitated force sensor to detect the exotic interaction at the millimeter scale. Improved constraints for the coupling $g_A^e g_V^N$ are established, within the force range spanning from 0.15 mm to 1.5 mm. And the constraint $|g_A^e g_V^N| \leq 4.39 \times10^{-26}$ at $\lambda$ = 0.5 mm at the $95 \%$ confidence level, significantly surpasses previous results by more than three orders of magnitude. The diamagnetic levitation force measurement system developed here can also be repurposed to probe other exotic spin-dependent interactions, such as the exotic spin-spin interaction. This system provides a platform for the exploration of dark matter.

\end{abstract}

\maketitle
\textit{Introduction.}\textbf{---}Various theories have predicted the existence of exotic interactions beyond the standard model~\cite{Cong2024}. Searching for exotic spin-dependent interactions, particularly at the low-energy frontier, in precise experiments has attracted broad interest in recent years ~\cite{PhysRevD.78.092006,Piegsa2012,Hunter2013,Heckel2013,Kotler2015,PhysRevLett.115.201801,PhysRevD.95.075014,Rong2018,NatCommun.10.2245,Ding2020,PhysRevLett.125.201802,sciadv.abi9535,PhysRevLett.127.010501,NationalScienceReview.10.nwac262,Wang2022,PhysRevLett.130.133202,Huang2024}.  First proposed by Moody and Wilczek ~\cite{PhysRevD.30.130}, the exotic spin-dependent interactions have since been expanded to encompass a broader spectrum of possibilities up to sixteen types ~\cite{JournalofHighEnergyPhysics.2006.005,PhysRevA.99.022113}. The exotic interactions between fermions are understood to arise from the exchange of light bosons, such as axions ~\cite{PhysRevLett.38.1440,RevModPhys.82.557}, dark photons ~\cite{AN2015331},  new $Z'$ bosons ~\cite{PhysRevLett.129.161801} and conformal sectors~\cite{Costantino2020}, which are proposed to explain the nature of dark matter ~\cite{Science.347(6226).1100-1102}. Specifically, there is a form of exotic interaction between electron spins and unpolarized nucleons, whose potential energy can be expressed as: 
\begin{equation}
	\label{displacement1}
	V=g_A^e g_V^N  \frac{\hbar} {4 \pi r}(\hat{\boldsymbol{\sigma}}\cdot\boldsymbol{v} )e^{- \frac{r}{\lambda}}.
\end{equation}
Here, $\hbar$ is Planck's constant, $\hat{\boldsymbol{\sigma}}$ is the spin unit vector of the electron, $\boldsymbol{v}$ and $r$ are respectively the relative velocity and the distance between the electron and the nucleon, $\lambda=\hbar/m_b c$ is the interaction range with $m_b$ corresponding to the mass of light boson, $g_A^e$ is the axial-vector coupling constant, and $g_V^N$ is the nucleon-vector coupling constant.

\begin{figure*}
	\centering
        \label{exp-sys1}\hypertarget{exp-sys1}{}
	\includegraphics[width=2\columnwidth]{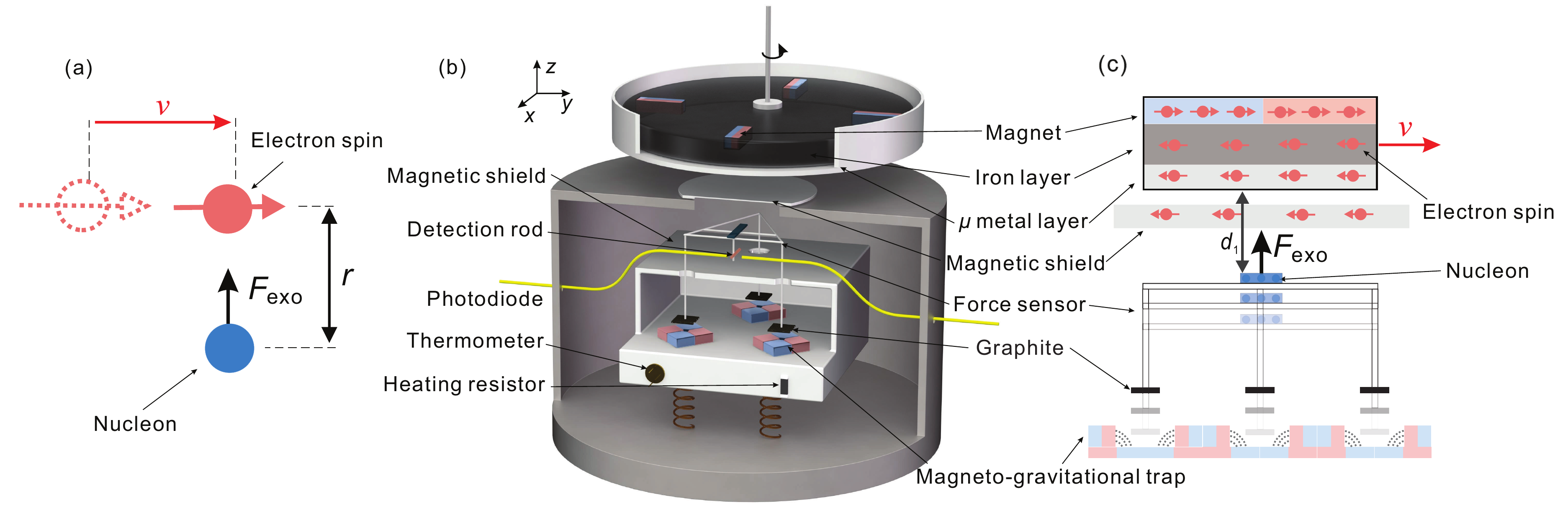}
	\caption{(Color online). Schematic of experimental principle and setup. (a) The exotic interaction occurs between a moving electron spin and a nucleon. (b) The exotic interaction is generated by four magnets, magnetized $\mu$-metal layer and magnetized iron layer. Four magnets are evenly distributed around the axis of rotation. The force sensor is composed of sapphire, frame, detection rod, and graphite segments. The graphite segments provide levitation force for the force sensor. The sapphire serves as the source of nucleons in this experiment. The detection rod partially blocks the light from the optical fiber to reflect the motion of the force sensor. The horizontal rotation of the rotating plate generates a relative velocity between the force sensor and the spin source. (c) The red arrow indicates the electron spins inside the magnet, iron layer, and $\mu$-metal layer, while the blue sphere represents the unpolarized nucleons in the force sensor. The red and blue colors on the magnets indicate the N pole and S pole, respectively. The anticipated exotic interaction occurs along the $z$-direction, resonantly exciting the translational $z$-mode of the force sensor. $d_1=1.57~\textrm{mm}$ denotes the distance between the force sensor and the $\mu$-metal layer. 
	}

\end{figure*}

Extensive experimental efforts have been dedicated to quantifying the velocity- and spin-dependent interaction, such as nitrogen-vacancy (NV) centers in diamonds \cite{PhysRevLett.127.010501,NationalScienceReview.10.nwac262} and spin exchange relaxation free (SERF) magnetometers \cite{NatCommun.10.2245,sciadv.abi9535}. These sophisticated instruments employ spin sensors to detect the effective magnetic field produced by the moving nucleons. Another method utilizes force sensors, such as the torsion balance \cite{PhysRevD.78.092006}, to directly measure the exotic force between electrons and nucleons. Torsion balances are renowned for their exquisite sensitivity in force detection. However, the detection distances are typically over 100 mm for measuring velocity and spin dependent exotic interaction. To date, there have been no reports of detecting velocity- and spin-dependent exotic interaction at the millimeter scale using force sensors.

In recent years, levitated mechanical systems on scales ranging from millimeters to nanometers have garnered significant interest in the field of precision measurement. Force measurements based on various forms of levitation, such as diamagnetic \cite{NewJPhys.20.063028,PhysRevApplied.15.024061,ApplPhysLett.12.124002}, superconducting \cite{ApplPhysLett.115.224101,PhysRevLett.131.043603,Sci.Adv.eadk2949}, electrical \cite{PhysRevLett.114.123602,PhysRevLett.132.133602}, and optical levitation \cite{NatPhy.9.1745-2481, NatNanotechnol.9.425,science.aba3993,PhysRevLett.128.111101} can achieve extremely high sensitivity. Currently, levitated force sensors are emerging as novel and effective systems for the detection of dark matter \cite{Carney_2021,Moore_2021}. Several promising proposals have been put forward \cite{Li_2023,PhysRevD.109.055024}. 
In particular, the diamagnetically levitated force sensor has been demonstrated to be an advantageous system for force measurements at sub-millimeter scales \cite{PhysRevApplied.16.L011003,NatPhys.18.1181–1185}, which may provide a new platform for detecting exotic interactions at such scales.

In this Letter, utilizing a diamagnetic levitated force sensor, we have developed a force measurement system to search for the exotic interaction between electrons and unpolarized nucleons at the millimeter range. We have constructed a rotating spin source with azimuthally varying electron polarization that provides a periodic force signal on the force sensor.  
The geometry of the spin source and force sensor is optimized to maximize the expected exotic interaction at the millimeter range. Additionally, it is crafted to minimize the intrusion of magnetic noise on the expected force signal, ensuring a clear and accurate measurement of the spin-dependent interactions. Consequently, this experiment has established a constraint on the coupling parameter, with $g_A^e g_V^N \leq 4.39 \times10^{-26}$ at $\lambda$ = 0.5 mm.

\textit{Experimental principle and system.}\textbf{---} In Fig.\ \hyperlink{exp-sys1}{1(a)}, a red sphere represents an electron, with the arrow indicating its direction of spin polarization, and a blue sphere represents a nucleon. The exotic interaction force is derived from Eq.\ \eqref{displacement1} as:
\begin{equation}
	\label{displacement2}
	\boldsymbol{F}=g_A^e g_V^N \frac{\hbar}{4 \pi}(\frac{1}{\lambda r}+\frac{1}{r^2})(\hat{\boldsymbol{\sigma}}\cdot\boldsymbol{v} )\hat{\boldsymbol{r}}e^{- \frac{r}{\lambda}}.
\end{equation} 
Here, $\hat{\boldsymbol{r}} = \boldsymbol{r}/|\boldsymbol{r}| $ is an unit vector that represents the relative position between the two particles. Likewise, when there is a relative velocity between a source mass containing electron spins and a force sensor composed of nucleons, there will be an exotic interaction between them. Based on this principle, we have established our force measurement system to search for the exotic interaction.

 Figure\ \hyperlink{exp-sys1}{1(b)} shows the schematic of the experiment setup \cite{See-supplemental-materials-for-details-of-the-data}. The force sensor, placed on a vibration isolation stage in vacuum, contains the nucleons to detect the exotic interaction. The force sensor is composed of four components: sapphire, frame, detection rod, and graphite segments \cite{NatPhys.18.1181–1185}. The sapphire component acts as a source of nucleons. The detection rod blocks a portion of the light emitted by the laser, with the remaining light being captured by a photodiode, thus recording the position of the force sensor. The frame is made of fiberglass of diameter 80 $\mu$m. The graphite segments levitate above the magneto-gravitational traps due to diamagnetism. To minimize the impact of the magnets in the magneto-gravitational traps on the experiment, we added a magnetic shield layer to the magnets.

The spin source, depicted in Fig.\ \hyperlink{exp-sys1}{1(b)} as the rotating plate, consists of three parts: magnets, iron layer, and $\mu$-metal layer. Four magnets, each measuring 8 mm by 2.5 mm by 2.5 mm, are symmetrically distributed around the axis of rotation. The iron layer is a cylinder with a diameter of 58 mm and a thickness of 3.5 mm. Four recesses, each with the same volume as the magnets, are created on the upper surface of the iron layer, allowing the magnets to fit perfectly and align with the upper surface of the iron layer. At the bottom of the spin source, a $\mu$-metal layer with a thickness of 0.7~mm and a diameter of 61~mm is placed. Additionally, a thin shell of $\mu$-metal is wrapped around the iron layer to further reduce the magnetic force signal.

There are polarized electron spins in the magnets, which contributes to the residual magnetism in the magnets. The iron layer and $\mu$-metal layer, as the soft magnetic materials, would be magnetized by the magnets. As a result, the magnetic fields of the magnets are largely shielded by the iron layer and $\mu$-metal layer. However, at different $z$-coordinates, the polarization distribution of the electron spins in the iron layer and the $\mu$-metal layer is different. For example, at bottom of the $\mu$-metal layer, the polarization orientations reverse along the azimuthal direction. This reversal is particularly significant because it provides the vast majority of contributions to the exotic interactions acting on the force sensor, owing to the close spacing. More details about the magnetic field and spin density distribution in the spin source can be found in Supplemental Materials \cite{See-supplemental-materials-for-details-of-the-data}. Finally, we have constructed a rotating spin source with closed magnetic circuits as well as electron spin density varying along the azimuthal angle.

The spin source rotates above the force sensor with a relative velocity between the force sensor, expressed as the relative velocity $\boldsymbol{v}$ in Eq. (\ref{displacement1}). A periodic exotic interaction force is anticipated to occur between the nucleons in the sapphire component of the force sensor and the polarized electron spins in the magnets, iron layer, and $\mu$-metal layer. In the experiment, we fine-tuned the rotating velocity of the plate to set the frequency of the periodic force signal at resonance with the $z$-mode of the force sensor \cite{See-supplemental-materials-for-details-of-the-data}. Consequently, the force sensor would be driven to its maximum amplitude by the anticipated exotic interaction, which is measured by recording the displacement of the force sensor. 

Apart from the anticipated periodic exotic interaction force, the rotating plate may also produce additional magnetic and electrostatic forces, as well as  Newtonian gravity.  The magnetic forces originate from the diamagnetism of the force sensor and constitute the primary confounding signal in this experiment. By leveraging the different distance dependencies of the exotic interaction and the conventional magnetic forces, we optimized the geometry of the spin source to maximize the exotic interaction force, primarily contributed by the electron spins in the $\mu$-metal layer, while effectively suppressing the magnetic forces. In order to further reduce the magnetic and electrostatic forces, we added a conducting magnetic shield layer between the rotating plate and the force sensor. While there are polarized electron spins in the magnetic shield due to magnetization from the magnets above, they do not contribute to the exotic interaction force signal because there is no oriented relative velocity between these electron spins and the force sensor. In addition, we have suppressed Newtonian gravity to a negligible magnitude by embedding the magnets in the iron layer with similar mass density.

\begin{figure}	
	\label{exp-sys2}\hypertarget{exp-sys2}{} 
	
	\includegraphics[width=0.96\columnwidth]{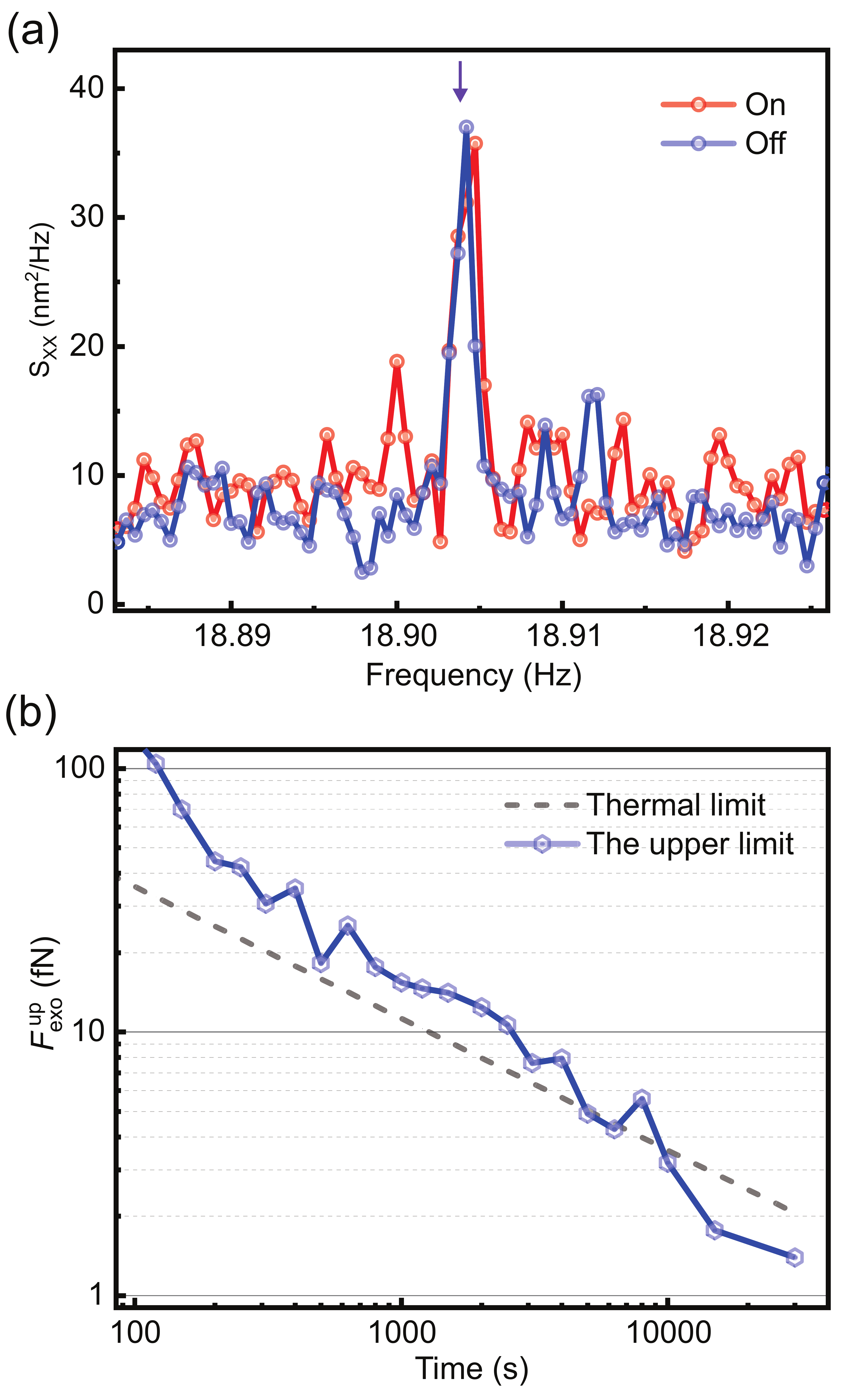}
	
	\caption{(Color online). Experimental results. (a) The power spectral density of the displacement of the force sensor $S_{\rm XX}$. We calculated the power spectral density 15 times over a period of 1900 seconds to obtain its average. The red and blue curves represent the power spectral density of the force sensor when the motor is turned on and off, respectively. The drive frequency of the spin source $\omega_{\rm dri}$$/2\pi$ matches the resonant frequency of the force sensor $\omega_0$/2$\pi$ ($\omega_{\rm dri}$/2$\pi$ = $\omega_0$/2$\pi$), as indicated by the purple arrow. (b) The blue line shows the upper limit of the exotic interaction at the 95\% confidence level, as a function of measurement time. The gray dashed line indicates the theoretical limit set by Brownian thermal noise.
	}
	
\end{figure}

As shown in Fig.\ \hyperlink{exp-sys1}{1(c)}, the electron spins of the rotating plate originate from three distinct components: the $\mu$-metal layer, iron layer, and  magnets \cite{PhysRevLett.115.201801,PhysRevD.95.075014,PhysRevLett.125.201802,PhysRevLett.130.133202}. The electron spin density within the magnets is determined by their residual magnetism. For the iron layer and the $\mu$-metal layer, the electron spin density is induced through magnetization by the magnets. As depicted in Fig.\ \hyperlink{exp-sys1}{1(c)}, the red spheres, each with an arrow, represent the electron spins. It is evident that the direction of electron spins within the magnet is anti-parallel to that within the iron layer and $\mu$-metal layer. Therefore, the resulting exotic interaction force is also in the opposite direction.

The total exotic interaction force between the rotating plate containing electron spins and the force sensor with nucleons is
\begin{equation}
	\label{displacement3}
	\boldsymbol{F}_{\rm exo}=\sum\limits_i \iint \boldsymbol{F} \rho_i  \rho_{\rm N} {\rm d} V_{i} {\rm d} V_{\rm N} .
\end{equation}
Here, $\boldsymbol{F}$ is the exotic interaction force between a nucleon and an electron spin as defined by Eq. (\ref{displacement2}). The index $i$ sequentially pertains to the components: magnets, iron layer, and $\mu$-metal layer. $V_{i}$ denotes the volume of each component respectively, $\rho_i$ denotes the density of each material, $V_{\rm N}$ is the volume of the force sensor, and $\rho_{\rm N}$ is the nucleon density of the force sensor.

The electron spin density of $\mu$-metal layer and iron layer can be calculated using the formula $\rho={f}M/{\mu_{\textrm{B}}}$, where $M$ represents the magnetization, $f$ is the spin contribution factor, and $\mu_{\textrm{B}}$ is the Bohr magneton. For $\mu$-metal layer, $f_\mu$ = 0.936 \cite{PhysRevB.63.094417}, and for iron layer,  $f_{\rm iron}$ = 0.957 \cite{AppliedPhysicsLetters.19.82-84}. The electron spin density for the magnets (${\rm Nd_2Fe_{14}B}$) can be calculated in the similar way using the formula $\rho=f_{\rm mag}M/{\mu_{\textrm{B}}}$, where $f_{\rm mag}$ = 0.982 \cite{PhysRevLett.85.429,PhysRevB.81.214408,PhysRevMaterials.3.064402}. The impact on the spin contribution factor $f_{\rm mag}$ from Nd, Fe, and B elements need to be considered separately. The magnetization of the magnets is determined by our measurement of the residual magnetism of the magnets, while the magnetization of $\mu$-metal layer and iron layer is calculated through finite element simulation. And we have conducted a conformation experiment using a magnet and a piece of $\mu$-metal layer to verify the reasonableness of the electron spin density in $\mu$-metal layer and iron layer simulated using finite element simulation in our experiment \cite{See-supplemental-materials-for-details-of-the-data}.

\textit{Data analysis.}\textbf{---}In this experiment, the force sensor was subjected to various forces in addition to the exotic interaction force of interest. These forces include  magnetic forces, electrostatic forces, mechanical vibrations, and Brownian thermal noise. The motion equation of the force sensor can be expressed as follows:
\begin{equation}
	\label{displacement}
	m\ddot X(t)+m\gamma\dot X(t) +m\omega_0^2X(t)=F_{\rm dri}(t)+F_ {\rm th}(t),
\end{equation}
where $m$ is the mass of force sensor, $\gamma$/2$\pi$ is the mechanical dissipation rate, $\omega_0$/2$\pi$ is the resonant frequency. $F_{\rm th}(t)$ is the Brownian thermal noise. $F_{\rm dri}(t)$ represents the periodic drive forces from the rotating plate that may include exotic interaction force, magnetic force, and electrostatic force. The effects of magnetic and electrostatic forces are effectively eliminated below the sensitivity of the force sensor. And the influence of mechanical vibrations is suppressed below the Brownian thermal noise, by placing the force sensor on a vibration isolation stage. Additionally, we conducted a calibration by Newtonian gravity measurement, the results of which demonstrate that the force sensor is thermally limited \cite{See-supplemental-materials-for-details-of-the-data}. 

During the experiment, we conducted two independent measurements to demonstrate the absence of exotic interaction within the range of our force detection. In the first measurement, we fine-tuned the rotating velocity of the spin source to set the frequency of the periodic exotic interaction at the resonance with the $z$-mode of the force sensor. We measured the force sensor's displacement in this `on' state, denoted as $X_1(t)$. In the second measurement, serving as a control, we turned off the motor, and denoted the displacement in this `off' state as $X_2(t)$.
After obtaining $X(t)$ for the two cases, the corresponding displacement response $X(\omega_{\rm dri})$ of the force sensor is calculated with
$X(\omega_{\rm dri})=\frac{1}{T}  \left|\int X(\tau) e^{i\omega_{\rm dri} \tau}d\tau\right|$, where $T$ is the measurement time. If there is such an exotic interaction drive force, it can be obtained with $F_{\rm dri}(\omega_{\rm dri})=mX(\omega_{\rm dri})/\chi(\omega_{\rm dri})$, where $\chi(\omega)=1/(\omega_0^2-\omega^2+i\omega \gamma)$ is the mechanical response of the force sensor.

Figure \hyperlink{exp-sys2}{2(a)} shows the power spectrum density of $X(t)$, defined as $S_{XX} = 2T\left<X^2(\omega)\right>$. The blue curve represents the power spectral density of $X_2(t)$ when the motor is off, contributed solely by the Brownian thermal noise.  Correspondingly, the red curve represents $X_1(t)$ when the motor is on, which may contain the signal of the exotic interaction force. After conducting continuous the measurements for a duration of 30,000 seconds, no positive force signal was observed within the margin of error. Figure \hyperlink{exp-sys2}{2(b)} shows the upper limit of the exotic force $F_{\rm limit}^{\rm up}$ at the 95$\%$ confidence level, varying as a function of the measurement time. The gray line shows the thermal noise limit. Thus, we establish the upper limit of the exotic interaction force as 1.40 fN, with 1 fN = $10^{-15}$ N.

\begin{figure}
	\label{exp-sys3}\hypertarget{exp-sys3}{}
	\includegraphics[width=0.96\columnwidth]{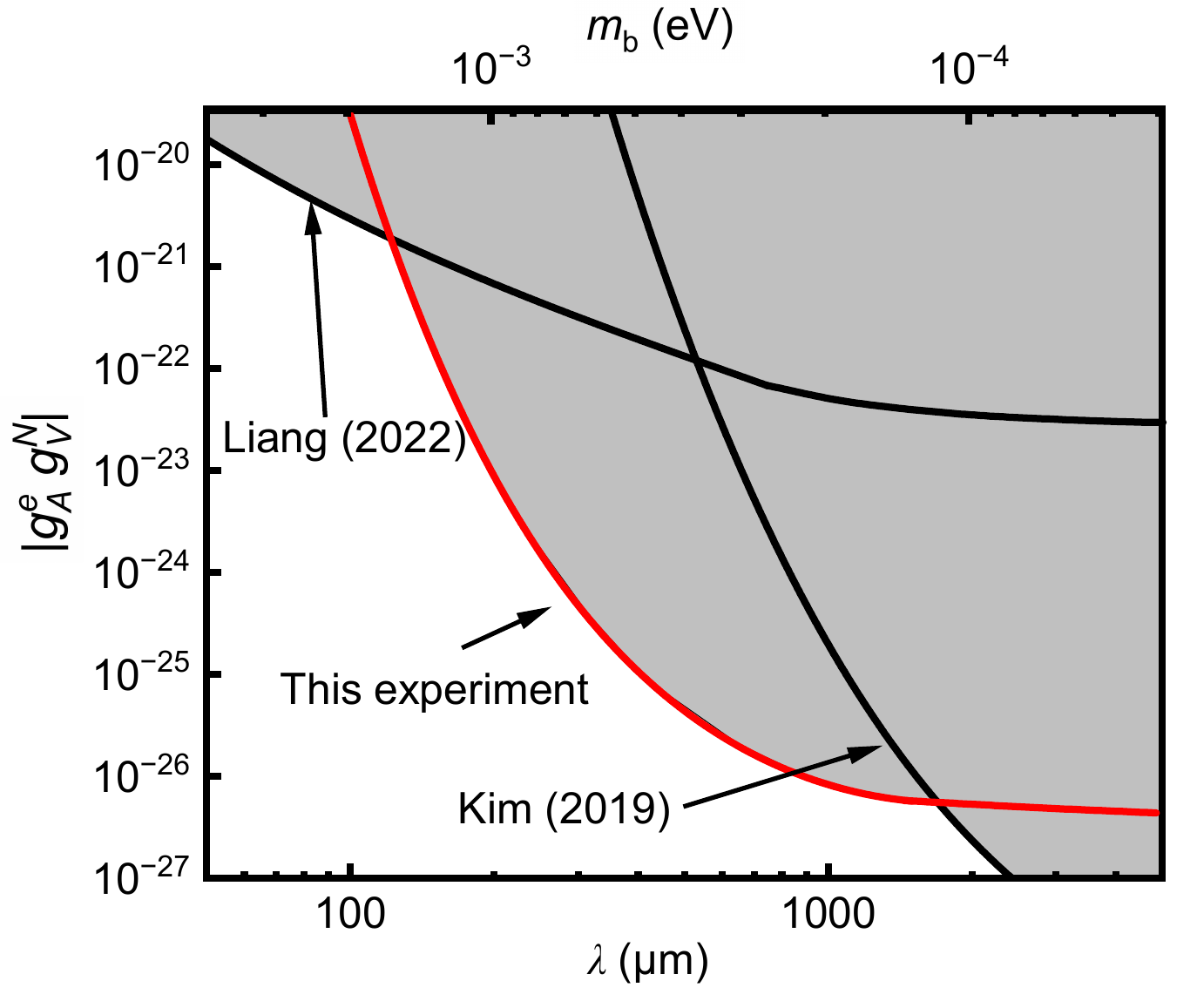}
	\caption{(Color online). The upper limit on the exotic spin and velocity dependent interaction $g_A^e g_V^N$ as a function of the force range $\lambda$. The black lines are the upper limits established by experiments in Refs. \cite{NationalScienceReview.10.nwac262,NatCommun.10.2245}. The red line is the upper bound obtained at the $95\%$ confidence level in this work, which establishes an improved laboratory bound in the force range from 0.15 mm to 1.5 mm. Notably, the constraint of the coupling factor $g_A^e g_V^N$ $\leq$ 4.39 $\times$ $10^{-26}$ at $\lambda=0.5$ mm is significantly improved by more than 3 orders of magnitude.}  
	
\end{figure}

Figure \hyperlink{exp-sys3}{3} plots the upper bounds on the spin and velocity dependent interaction established by this work, along with constraints from previous experiments. The gray shaded areas represent the parameter space excluded experimentally. In our experiment, the relative systematic error introduced by various factors is 31.0$\%$ \cite{See-supplemental-materials-for-details-of-the-data}. The red solid line represents the constraint established by this experiment at the $95\%$ confidence level, taking the upper limit of the exotic interaction force as $1.40$ fN. Kim et al.  \cite{NatCommun.10.2245} imposed the strongest constraints for the force range $\lambda$ $\textgreater$ 1.5 mm using spin exchange relaxation free magnetometer, and Liang et al.~\cite{NationalScienceReview.10.nwac262} established upper bounds for the force ranges $\lambda$ $\textless$ 0.15 mm using nitrogen-vacancy centers. Our experiment addresses the intermediate force range of 0.15 mm to 1.5 mm, providing the best experimental bounds. In particular, at $\lambda$ = 0.5 mm, the constraint on experimental indicator $g_A^e g_V^N$  has been improved by more than three orders of magnitude.

\textit{Discussion and summary.}\textbf{---}We have developed a method to search for the velocity and spin dependent exotic interaction, utilizing a diamagnetically levitated force sensor. We have demonstrated the effectiveness of our system at the millimeter scale, showing that the experimental indicator $g_A^e g_V^N$ at $\lambda$ = 0.5 mm has been improved by more than three orders of magnitude over previous experiments.

In the future, we envision several methods to further improve the sensitivity of our force sensor. Firstly, we can greatly reduce the impact of Brownian thermal noise by reducing the temperature to 3 K. Secondly, the current mechanical dissipation of the force sensor is primarily due to the eddy current in the graphite segments. We can efficiently reduce this eddy current dissipation using laser engraving technology \cite{PhysRevResearch.5.013030}. Thirdly, we can increase the mass of the force sensor to increase the number of nucleons within the force sensor. Based on the above methods, we expect that $g_A^e g_V^N$ at $\lambda$ = 0.5 mm can be further improved to the order of $10^{-28}$.

Furthermore, our experimental setup is not only capable of measuring velocity and spin-dependent exotic interactions. It is also proficient in examining other exotic spin-dependent phenomena including, for instance, exotic spin-spin interaction between a pair of polarized electrons. This versatility highlights the potential of our experimental setup to make a broad contribution to the field of fundamental physics research.
\hspace*{\fill}

\begin{acknowledgments}
    We thank Min Jiang and Gao Yu for helpful advices. This work was supported by the National Natural Science Foundation of China (Grants No.\ T2388102, No.\ 12150011, No.\ 12075115, No.\ 62071118) and the Primary Research $\&$ Development Plan of Jiangsu Province (Grant No. BE2021004-2), and the Natural Science Foundation of Jiangsu Province (Grant No. BK20241255).

    K. T., Y. S., R. L. and L. W. contributed equally to this work.
\end{acknowledgments}

\end{document}